# "Does it Chug?"
# Towards a Data-Driven Understanding of Guitar Tone Description

**Pratik Sutar**[1]
sutarpratik2012@gmail.com

**Jason Naradowsky**[1,2]
narad@is.s.u-tokyo.ac.jp

**Yusuke Miyao**[1]
yusuke@is.s.u-tokyo.ac.jp

[1]The University of Tokyo, [2]Square-Enix

## Abstract

Natural language is commonly used to describe instrument timbre, such as a "warm" or "heavy" sound. As these descriptors are based on human perception, there can be disagreement over which acoustic features correspond to a given adjective. In this work, we pursue a data-driven approach to further our understanding of such adjectives in the context of guitar tone. Our main contribution is a dataset of timbre adjectives, constructed by processing single clips of instrument audio to produce varied timbres through adjustments in EQ and effects such as distortion. Adjective annotations are obtained for each clip by crowdsourcing experts to complete a pairwise comparison and a labeling task. We examine the dataset and reveal correlations between adjective ratings and highlight instances where the data contradicts prevailing theories on spectral features and timbral adjectives, suggesting a need for a more nuanced, data-driven understanding of timbre.

## 1 Introduction

The study of music, whether through performance or appreciation, takes us on an ever-deepening journey to understand its many complexities. Among these complexities is the characteristic sound of the instruments, a property known as *timbre*. Within circles of musicians and music aficionados, unique vocabularies emerge to help articulate the subtle and intricate characteristics of instrument sounds. While common terms like *bright* or *dark* might resonate with a wide audience, others such as *dry*, *fat*, *lush*, and *round* introduce further nuance and intricacy. These terms, rich in nuance, aim to bridge the gap between the physical experience of sound and its emotional impact. However, a challenge arises in establishing a shared understanding of these descriptors: What defines the qualities that constitute a dry or fat sound? And more importantly, how can we navigate the subjective nature of sound perception to agree on what these terms truly signify?

To better understand how timbre adjectives are invented, and how online communities reach a consensus on their meanings, we construct a new dataset of aligned audio clips with varying timbres, annotated with adjective labels and pairwise comparison among the clips. Our study focuses on a single instrument: the electric guitar, motivated by (a) its extensive use across a broad spectrum of contemporary musical genres, (b) the presence of a rich community of online discussion forums for guitar enthusiasts that have given rise to many unique timbral adjectives (what does it mean to *chug*? What is a *brown* sound?), and (c) while the instrument inherently contributes certain timbral characteristics, it is predominantly the application of additional processing (effects, amplification) that shapes the sound into distinct timbres. This instrument choice enables us to apply different processing to a given guitar performance, creating many recordings where the timbre differs but the musical content remains constant. This approach allows us to isolate and study the effects of timbre independently from other factors. We release all code and dataset[1] to facilitate additional research and aid the development of language and music creation systems, such as prompt-based music generation (Agostinelli et al., 2023; Copet et al., 2024; Huang et al., 2023; Evans et al., 2024).

## 2 Related Works

The study of how we describe timbre, and the ways in which we create or borrow words to facilitate it, has a long history (Wake and Asahi, 1998; Porcello, 2004; Wallmark, 2019). Relevant to this work, it has been empirically found that experts, over a prolonged period of practice and exposure to various timbres, develop an ability to acutely distinguish between finer timbral variations and develop a sophisticated vocabulary to communicate them (Bernays and Traube, 2013). Studies support that experts rely more on timbral differences when communicating about novel sounds (Lemaitre et al., 2010), though the creative use of words is not limited to experts (Wake and Asahi, 1998).

---
[1]https://github.com/PratikStar/doesitchug

Also relevant to our work is how words are invented, or often borrowed from other contexts to fulfill a new role as a timbral descriptor. Among many studies on this topic, a recent study proposes a categorization of the origins of instrument timbre descriptors into seven classes (Wallmark, 2019). The descriptors in our proposed dataset are sufficiently diverse to have examples from each of these categories. Similar to our work, (Seetharaman and Pardo (2016)) use crowdsourcing to gather timbre annotations for recordings of audio effects, such as equalizers. Our work differs in that we focus on a variety of timbre for a single instrument and collect pairwise comparisons, and we construct our annotator pool of participants from online enthusiast communities.

A widely used quantitative method for studying perceptual qualities of timbre involves rating sound stimuli on a verbal scale. One approach is the Semantic Differential (SD) technique (Osgood, 1952), where each question involves rating adjective pairs that have opposing meanings, e.g., *dark-bright*, *smooth-rough*, etc. Due to the use of verbal scales, SD studies suffer from issues like polysemy and non-exact antonymy (*bright-dull* in (Pratt and Doak, 1976), *bright-dark* in (Alluri and Toiviainen, 2010)). A common solution is to use unipolar rating scales (Kendall and Carterette, 1993), which are bounded by an attribute (e.g. *soft*) and its negation (e.g. not *soft*). Of note to our study is that while many adjectives have obvious opposites, many others do not. We thus argue that the creation of larger data is necessary, in order to enable a data-driven understanding of these terms.

An alternative to verbal scales are dissimilarity studies, in which participants rate differences between pairs of sounds. Techniques like multi-dimensional scaling (MDS) are then used to produce a spatial arrangement where distances between points correspond to these dissimilarity ratings (Shepard, 1962). The latent dimensions of MDS can then be correlated with the physical characteristics of the sound (Peeters et al., 2011; Mcadams et al., 2014).

## 3 Dataset Creation

The dataset creation process involves three key steps, (1) collecting a comprehensive set of adjectives for describing tone from online communities, (2) generating audio recordings that encompass a broad range of timbres, and (3) annotating the recordings via crowdsourcing using an online interface. As our dataset consists of nuanced timbral distinctions within a singular instrument class, all data is of electric guitar recordings.

### 3.1 Collecting Timbre Descriptors

In this work, we aim to study how timbre and tone are discussed more informally, evolving as the need develops, in the niche or online communities discussing specific music tones, genres, or styles. Thus, we turn to those communities themselves to know which adjectives are commonly used outside the established literature. We begin by crawling the internet for articles discussing guitar timbre words, using keyword searches of the form "a(n) $x$ sound/tone" for a given adjective $x$. We also engage with these communities to gather additional suggestions. This process resulted in a set of 110 adjectives, which are presented in the appendix A.

### 3.2 Creating the Audio Files

To study a diverse set of timbral descriptors, it is necessary to generate a diverse set of instrument audio recordings such that they could foreseeably be described using a wide range of the adjectives gathered in the preceding step. We approach this problem using a two-step process, first generating unprocessed guitar sounds in a variety of genres (diverse content), and then processing them with different signal processing chains to yield a variety of sounds (diverse timbre).

First, we record a series of unprocessed signals, also known as direct input (DI), from an electric guitar without any sound shaping. We hypothesize that some timbral descriptions may only apply to specific genres or styles of playing. For instance, very percussive and fast rhythm playing is unlikely to be described as *chimey* regardless of the instrument timbre. Therefore, to capture a variety of playing styles, we collect a number of recordings from three different guitar players, one amateur and two professional.

We manually sample segments from these recordings, aiming to select short segments representing a diverse set of styles and dynamics. The final set of DI contained 12 recording segments with content ranging from slow arpeggios, simple chords, aggressive-style rhythm playing, and fast soloing. Each segment is approximately 10 seconds in length, 44.1kHz monaural audio.

We then process each DI using a different FX chain to achieve a diverse set of timbres. For this, we use a commercial plugin (*Helix Native*) which emulates various effects, amplifiers, and cabinets. To ensure that these chains generate desirable sound, we utilize the included presets, which are specific parameter settings designed by manufacturers of audio plugins, artists, or other users to achieve a specific tone of interest. We process each of the 12 DI clips using the 80 preset effects to produce 960 audio samples. A complete list of the presets can be found in the appendix B. The processing of audio signals is performed using REAPER.

### 3.3 Annotation Interface Design

We design a web interface for collecting annotations, in which we collect three types of annotations.

#### 3.3.1 Pairwise annotations

The annotator is presented with two samples, A and B, in random order. For a given adjective X, the annotator is asked to choose: (1) A is more $X$ than B, (2) A is less $X$ than B, (3) Both audio samples are equally $X$, or (4) to skip the question.

Each audio sample A and B is based on the same DI recording, and thus their musical content is identical. This allows the user to focus solely on the differences in timbre, and to minimize the confounding aspects of other acoustic factors, such as pitch and loudness, which have been noted to affect the perception of timbre (Melara and Marks, 1990; McAdams and Goodchild, 2017).

The benefit of the ranked comparison is that it allows us to gather data about very precise timbral relationships, e.g., in situations where the overall sound of timbre A vs. B is presumably much closer than that of previous work, where such clips would represent different instruments entirely. Second, ranking directly supports important practical use cases, such as "In which of these songs is the sound of guitar more $X$?".

#### 3.3.2 Label annotations

Pairwise rank comparison can be an extremely informative annotation, but because we must arrange comparisons randomly in order to avoid imparting any bias to the study, some ranked comparisons will be less useful and irrelevant. The ternary nature of our ranked comparison (an (A, B, $X$) tuple) may also lead to sparsity. In order to counteract this and ensure more information-per-recording, we also collect label annotations. After the annotator has made a ranked comparison, the annotator is asked to select any adjectives from the adjective list that may apply to the selected clip of the pairwise annotation.

#### 3.3.3 Custom Annotations

A final source of annotations is an open text field, where annotators may enter any other adjectives that apply to the selected clip and are not contained in the adjective list. These adjectives aren't included in the annotation list but are retained in the dataset for future research.

### 3.4 Collecting Annotations

We seek to understand more nuanced descriptions of tone that arise in online communities under the need to describe increasingly specific timbral qualities. By the very nature of the study, a pool of general annotators (like those commonly hired via *Mechanical Turk*) is not appropriate for the study, as they lack the expertise and experience in discussing these sounds. Instead, we enlist volunteers from online guitar and music enthusiast communities by incentivizing participation using an online raffle system. In total, we collect 2038 annotations from 38 participants. In addition to timbral annotations, we also record participant information, such as where they heard about the study, and how many years of experience they have playing the guitar. Notably, 87% of our annotators have more than 10 years of experience playing guitar.

### 3.5 Unifying Annotations

As we collect multiple types of annotation on the level of individual clips, we present a method to unify the annotations and provide a single score for each clip-adjective combination (which can then be averaged over clips to provide a score between any preset/timbre and adjective). For pairwise comparisons, models like Bradley–Terry (Bradley and Terry, 1952) can be used, however, as we also include multi-label annotations on clips, we instead present a simple graph-based algorithm that combines the two types of annotations for its potential future use.

For every adjective in the label annotations, we add a constant $\phi$ to the presets labeled with the adjective, representing a single "unit" of adjective-preset correlation. Working with these ratings, we utilize pairwise annotations to discover and enhance the *greater than* or *less than* relationships among the data. For every adjective, we find the set of presets, $\{H\}$, with the highest label annotation score. From the pairwise comparison data, we then find the relationships where $A$ is rated less than $B$ and $A \in \{H\}$. In alignment with such pairwise comparisons, we adjust the score of $B$ to be greater than $A$ by a constant, $\phi$. We then infer scores lower than the lowest label annotation score. We repeat this inference process until no new higher or lower preset is found. In the case of ties, we prioritize the pairwise annotation data over the label annotations. We release these scores with the dataset.

## 4 Analysis

### 4.1 Presets By Adjectives

The table 1 shows presets corresponding most to a sampling of adjectives. Evaluating the correctness of a dataset of this type is difficult, as by its very nature there is no gold standard to refer to. However, we find many of the highly correlated presets correspond well to known descriptions of the sounds they are modeled on. For instance, *07B Line6 Litigator*, which is ranked in the dataset as being most correlated to *warm*, is based on a Dumble Overdrive amplifier, which is expertly described as having a "*very open and uncompressed feel, overdrive without fuzz, warm sustaining cleans, and of course that saxophone-like midrange and sing that these amps are famous for*"[2]. We encourage the reader to listen to the clips for a better understanding of the extent to which these presets relate to these adjectives.

### 4.2 Novel Findings

Existing work, utilizing unaligned audio of different instruments, has identified spectral features that correlate with the perception of acoustic properties, which we describe using timbral adjectives (Schubert and Wolfe, 2006). The annotations of our dataset allow us to revisit these claims and assess how well they agree with the crowdsourced consensus. We provide one case study on *brightness* and its relationship to the spectral centroid. We find that in pairs of clips which should be ranked as A > B with respect to existing theories, crowdsourced workers ranked them differently. Visualizations of these relationships are presented in the appendix C.2. We argue that these findings are evidence that further analysis into the acoustic causes of human perception of these properties is necessary.

---

[2]https://www.sebagosound.com/index.php?id=18

| Adjective | Most Relevant Preset |
|---:|:---|
| Abrasive | 18C THE BLUE AGAVE |
| Articulate | 11A BAS_Woody Blue |
| Bassy | 07D ANGL Meteor |
| Buzzy | 04A Jazz Rivet 120 |
| Clean | 09A DI |
| Distorted | 03C Brit 2204 |
| Twangy | 04A Jazz Rivet 120 |
| Warm | 07B Line 6 Litigator |

Table 1: The most relevant preset for various adjectives, as calculated by the graph-based unification algorithm.

### 4.3 Inter Annotator Agreement

As we aim to compare a variety of audio samples pairwise, across many adjectives, the number of possible comparisons is very high. And because annotators needed experience with the instrument, we're limited by how many possible data samples we can get, which naturally leads to sparsity and limits the ability to conduct inter-annotator agreement. However, amongst the 6 instances where we found multiple responses on the same annotation question, in only one case did the annotators disagree about the ranking of the clips.

## 5 Conclusions

In this work, we present a dataset that focuses on very fine-grained differences in timbre, isolating them from other factors by generating recordings of different timbres based on shared DIs, containing identical musical content. We find that human assessments sometimes differ from previously established correlations between coarse acoustic features and the perception of adjectives, supporting the need for a more nuanced understanding of acoustic correlates of these descriptors in the context of guitar music. Furthermore, this understanding will also yield practical improvements in prompt-based conditional audio generation, timbre-based music retrieval, and natural language interfaces for musical tools (Rosi, 2022).

# 6 Acknowledgements

We thank the guitarists Lorcan Ward and Ola Englund for granting permission to use their DI tracks. We also thank the many survey participants from The Gear Page, Sevenstring.org, Rig-talk, and The Sound of AI.

# References

Andrea Agostinelli, Timo I. Denk, Zalán Borsos, Jesse Engel, Mauro Verzetti, Antoine Caillon, Qingqing Huang, Aren Jansen, Adam Roberts, Marco Tagliasacchi, Matt Sharifi, Neil Zeghidour, and Christian Frank. 2023. Musiclm: Generating music from text. *Preprint*, arXiv:2301.11325.

Vinoo Alluri and Petri Toiviainen. 2010. Exploring perceptual and acoustical correlates of polyphonic timbre. *Music Perception - MUSIC PERCEPT*, 27:223–242.

Michel Bernays and Caroline Traube. 2013. *Expression of piano timbre: Verbal description and gestural control*, pages 205–222.

Ralph Allan Bradley and Milton E. Terry. 1952. Rank Analysis of Incomplete Block Designs: I. The Method of Paired Comparisons. *Biometrika*, 39(3/4):324–345.

Jade Copet, Felix Kreuk, Itai Gat, Tal Remez, David Kant, Gabriel Synnaeve, Yossi Adi, and Alexandre Défossez. 2024. Simple and controllable music generation. *Preprint*, arXiv:2306.05284.

Zach Evans, CJ Carr, Josiah Taylor, Scott H. Hawley, and Jordi Pons. 2024. Fast timing-conditioned latent audio diffusion. *Preprint*, arXiv:2402.04825.

Hermann L. F. Helmholtz. 1877. *On the Sensations of Tone as a Physiological Basis for the Theory of Music*. Original work in German, titled "*Die Lehre von den Tonempfindungen als physiologische Grundlage für die Theorie der Musik*" and published in 1863.

Qingqing Huang, Daniel S. Park, Tao Wang, Timo I. Denk, Andy Ly, Nanxin Chen, Zhengdong Zhang, Zhishuai Zhang, Jiahui Yu, Christian Frank, Jesse Engel, Quoc V. Le, William Chan, Zhifeng Chen, and Wei Han. 2023. Noise2music: Text-conditioned music generation with diffusion models. *Preprint*, arXiv:2302.03917.

Roger A. Kendall and Edward C. Carterette. 1993. Verbal Attributes of Simultaneous Wind Instrument Timbres: I. von Bismarck's Adjectives. *Music Perception: An Interdisciplinary Journal*, 10(4):445–467.

Guillaume Lemaitre, Olivier Houix, Nicolas Misdariis, and P. Susini. 2010. Listener Expertise and Sound Identification Influence the Categorization of Environmental Sounds. *Journal of experimental psychology. Applied*, 16:16–32.

Stephen Mcadams, Bruno Giordano, P. Susini, Geoffroy Peeters, and Vincent Rioux. 2014. A meta-analysis of acoustic correlates of timbre dimensions. volume 120, pages 3275–3276.

Stephen McAdams and Meghan Goodchild. 2017. Musical structure: Sound and Timbre. In *The Routledge Companion to Music Cognition*. Routledge.

Robert Melara and Lawrence Marks. 1990. Interaction among auditory dimensions: Timbre, pitch, and loudness. *"Perception & Psychophysics"*, 48:169–78.

Charles E. Osgood. 1952. The nature and measurement of meaning. *Psychological Bulletin*, 49:197–237.

Geoffroy Peeters, Bruno L. Giordano, Patrick Susini, Nicolas Misdariis, and Stephen McAdams. 2011. The Timbre Toolbox: Extracting audio descriptors from musical signals. *The Journal of the Acoustical Society of America*, 130(5):2902–2916. Publisher: Acoustical Society of America.

Thomas Porcello. 2004. Speaking of Sound: Language and the Professionalization of Sound-Recording Engineers. *Social Studies of Science*, 34(5):733–758.

R. L. Pratt and P. E. Doak. 1976. A subjective rating scale for timbre. *Journal of Sound and Vibration*, 45(3):317–328.

Victor Rosi. 2022. *The Metaphors of Sound: from Semantics to Acoustics - A Study of Brightness, Warmth, Roundness, and Roughness*. phdthesis, Sorbonne Université.

Emery Schubert and Joe Wolfe. 2006. Does timbral brightness scale with frequency and spectral centroid. *Acustica*, 92:820–.

Prem Seetharaman and Bryan Pardo. 2016. Audealize: Crowdsourced audio production tools. *Journal of the Audio Engineering Society*, 64:683–695.

Roger N. Shepard. 1962. The analysis of proximities: Multidimensional scaling with an unknown distance function. I. *Psychometrika*, 27(2):125–140.

Sanae Wake and Toshiyuki Asahi. 1998. Sound Retrieval with Intuitive Verbal Expressions. In *International Conference on Auditory Display '98*.

Zachary Wallmark. 2019. A corpus analysis of timbre semantics in orchestration treatises. *Psychology of Music*, 47:585–605.

# A Timbre Adjectives

| | | | | |
|---|---|---|---|---|
| Abrasive | Chug | Focused | Mellow | Shrill |
| Aggressive | Chunky | Full | Metallic | Sizzling |
| Airy | Clean | Fuzzy | Muddy | Smokey |
| Anemic | Clear | Glassy | Muffled | Smooth |
| Articulate | Compressed | Greasy | Muted | Soft |
| Artificial | Crisp | Grind | Nasal | Sparkly |
| Balanced | Crunchy | Gritty | Noisy | Sterile |
| Bassy | Crushing | Grotty | Open | Strained |
| Bell-like | Cutting | Grunting | Piercing | Strident |
| Big | Dark | Hairy | Punchy | Sweet |
| Biting | Delicate | Harsh | Pure | Thick |
| Bold | Detailed | Heavy | Raspy | Thin |
| Boomy | Dirty | Hissing | Raw | Throaty |
| Boxy | Distorted | Hollow | Refined | Thumping |
| Bright | Dry | Honky | Rich | Tight |
| Brilliant | Dull | Huge | Ringing | Tinny |
| Brittle | Dynamic | Icepicky | Round | Twangy |
| Brutal | Edgy | Jangly | Saturated | Velvety |
| Buzzy | Fat | Light | Scooped | Vibrant |
| Chewy | Fizzy | Liquidy | Searing | Vintage |
| Chimey | Flabby | Loose | Sharp | Vocal |
| Choked | Flat | Lush | Shimmery | Warm |

Table 2: The complete list of adjectives used in the study for pairwise comparison and label annotation.

| | | | | |
|---|---|---|---|---|
| Blunt | Defined | Nostalgic | Robotic | Wavey |
| Brittle | Defined | Plucky | Saturated | Wrapped |
| Chirping | Digital | Pointy | Scratchy | |
| Contained | Drive | Popping | Stuffy | |
| Crisp | Echoey | Pounding | Subdued | |
| Deep | Natural | Present | Telephone | |

Table 3: A list of custom adjectives collected from the annotators during the annotation process as described in Section 3.3.3.

# B List of Presets

| | | |
|---|---|---|
| 01A US Double Nrm | 01B Essex A30 | 01C Brit Plexi Jump |
| 01D Cali Rectifire | 02A US Deluxe Nrm | 02B A30 Fawn Nrm |
| 02C Revv Gen Purple | 02D Revv Gen Red | 03A Archetype Clean |
| 03B Matchstick Ch1 | 03C Brit 2204 | 03D Archetype Lead |
| 04A Jazz Rivet 120 | 04B Fullerton Brt | 04C Brit J45 Brt |
| 04D Solo Lead OD | 05A Placater Clean | 05B Interstate Zed |
| 05C Placater Dirty | 05D PV Panama | 06A Cali Texas Ch 1 |
| 06B Essex A15 | 06C Derailed Ingrid | 06D German Mahadeva |
| 07A WhoWatt 100 | 07B Line 6 Litigator | 07C Cartographer |
| 07D ANGL Meteor | 08A US Small Tweed | 08B Divided Duo |
| 08C Brit P75 Brt | 08D Line 6 Badonk | 09A DI |
| 09B BAS_SVT-4 Pro | 09C BAS_Cali Bass | 09D BAS_Aqua 51 |
| 10A BAS_Cougar 800 | 10B BAS_SVT Nrm | 10C BAS_Cali 400 Ch1 |
| 10D BAS_Del Sol 300 | 11A BAS_Woody Blue | 11B Trademark |
| 11C AUS Flood | 11D Justice Fo Y'all | 12A Lonely Hearts |
| 12B Pull Me Under | 12C Stone Cold Loco | 12D Plush Garden |
| 13A Cowboys from DFW | 13B G.O.A.T Rodeo | 13C BIG DUBB |
| 13D BIG VENUE DRIVE | 14A BUBBLE NEST | 14B DUSTED |
| 14C SUNRISE DRIVE | 14D GLISTEN | 15A WATERS IN HELL |
| 15B FAUX 7 STG CHUG | 15C RICHEESE | 15D RC REINCARNATION |
| 16A RIFFS AND BEARDS | 16B FELIX MARK IV | 16C FELIX JAZZ 120 |
| 16D FELIX DELUXE MOD | 17A FELIX ENGL | 17B SPOTLIGHTS |
| 17C BUMBLE ACOUSTIC | 17D BMBLFOOT PRINCE | 18A SHEEHAN PEARCE |
| 18B SHEEHAN SVT4PRO | 18C THE BLUE AGAVE | 18D BULB RHYTHM |
| 19A BULB LEAD | 19B BULB CLEAN | 19C BULB AMBIENT |
| 19D EMPTY GARBAGE | 20A ONLY GARBAGE | 20B GARBAGE BASS |
| 20C BILLY KASTODON | 20D THIS IS THE END | |

Table 4: A list of presets from Helix Native used for obtaining different timbres. See the guide for more detail.

## C Further Analysis

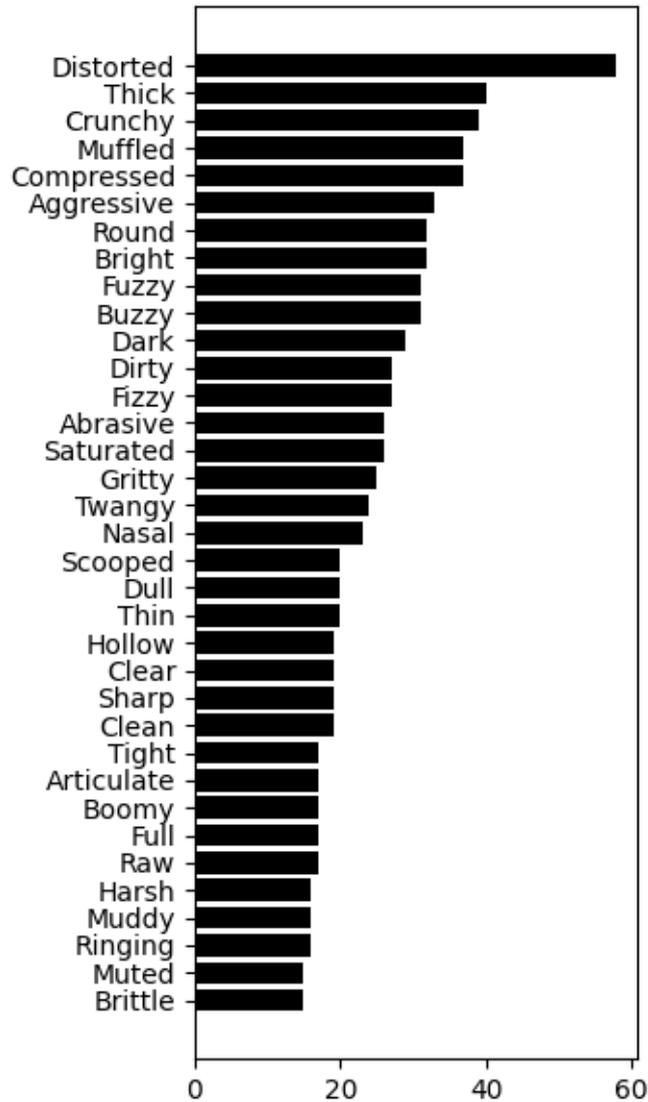

Figure 1: Frequencies of labels

### C.1 Label Frequencies

Figure 1 shows the most frequent 35 labels. Among the most annotated labels, we find a frequency of annotation of 20-40 times. Even among the top labels, we observe a good diversity in timbre, although there seems to be some skew towards heavier genres. This may be a bias in our dataset stemming from uniformly sampling the Helix presets, many of which are geared toward metal and rock genres. These labels cover all the categories proposed in the comprehensive taxonomy study (Wallmark, 2019), some examples from each of the categories are *Aggressive*, *Dull* from **Affect**; *Round*, *Full* from **Matter**; *Bright*, *Sharp* from **CMC**; *Boomy*, *Twangy* from **Mimesis**; *Muffled*, *Saturated* from **Action**; *Ringing*, *Muted* from **Acoustics**; and *Buzzy*, *Fizzy* from **Onomatopoeia**. This diversity underscores the richness and complexity of timbral descriptions in our dataset.

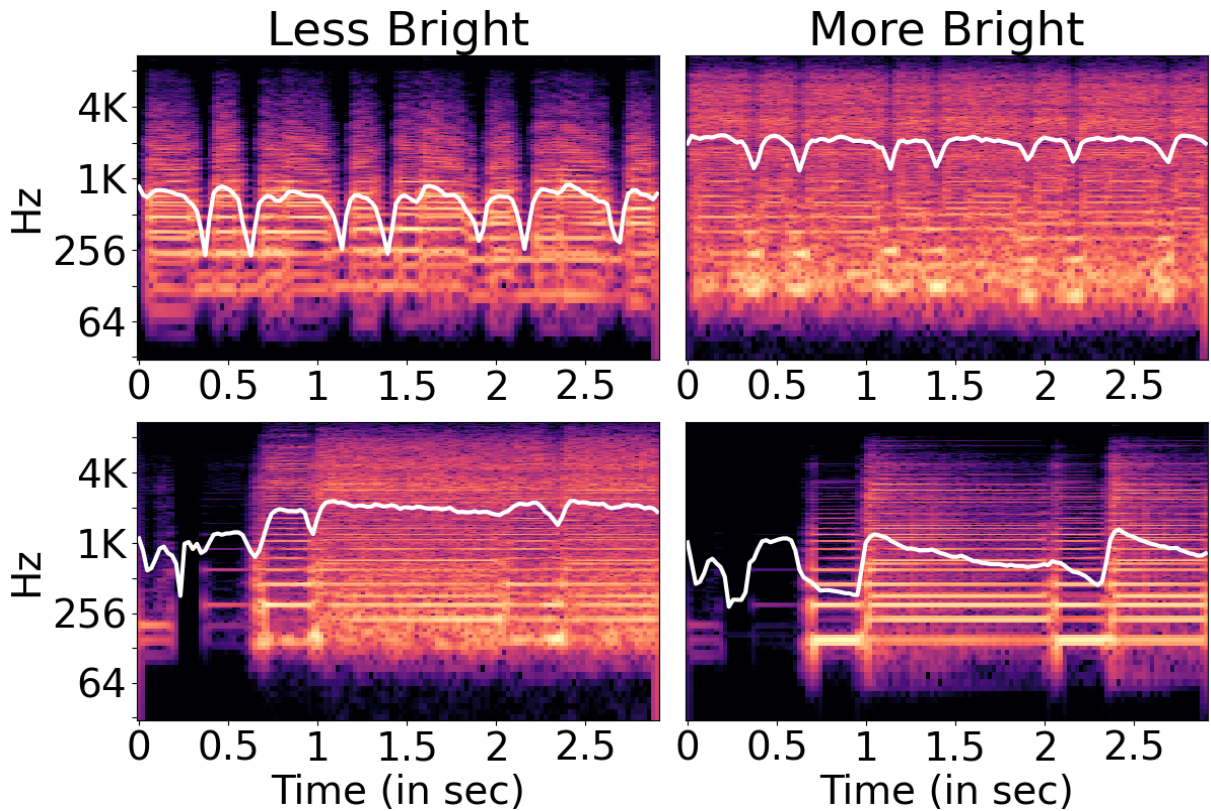

Figure 2: Each row represents one paired comparison. Audio on the right column is labeled **more** bright than the one on the left. In the top row, the pairwise annotation is consistent with the spectral centroid (shown in white), whereas it is not consistent with the centroid in the bottom row.

### C.2 Case Study: Spectral Centroid

"*Brightness*", which is a commonly studied timbral descriptor, dating back at least to (Helmholtz, 1877) and has more recently been correlated to the center of mass of the spectrum, often referred to as the spectral centroid (Schubert and Wolfe, 2006). While this result holds generally in our dataset, and recordings with higher spectral centroids are more likely to be labeled as "bright", we also observe many confounding factors. The rows of Figure 2 show spectrograms of pairwise comparison between two clips from our dataset where the left clip was annotated as less bright than the right one. In the top-row comparison, the spectrogram with the higher spectral centroid is indeed considered brighter, but in the second (bottom) comparison, the relationship does not hold.

Why is this the case? Although existing work on correlating spectral features to acoustic properties and adjectives provides a general approach, we hypothesize that other factors should be considered when correlating the acoustic feature to timbral adjectives. In the case of brightness, features like F0 and Harmonic-to-Noise ratio (HNR) may play a role (Rosi, 2022). However, the difficulty of understanding the interactions between these features and how they relate to brightness supports the notion that a more data-driven (or machine learning-driven approach) may be necessary.

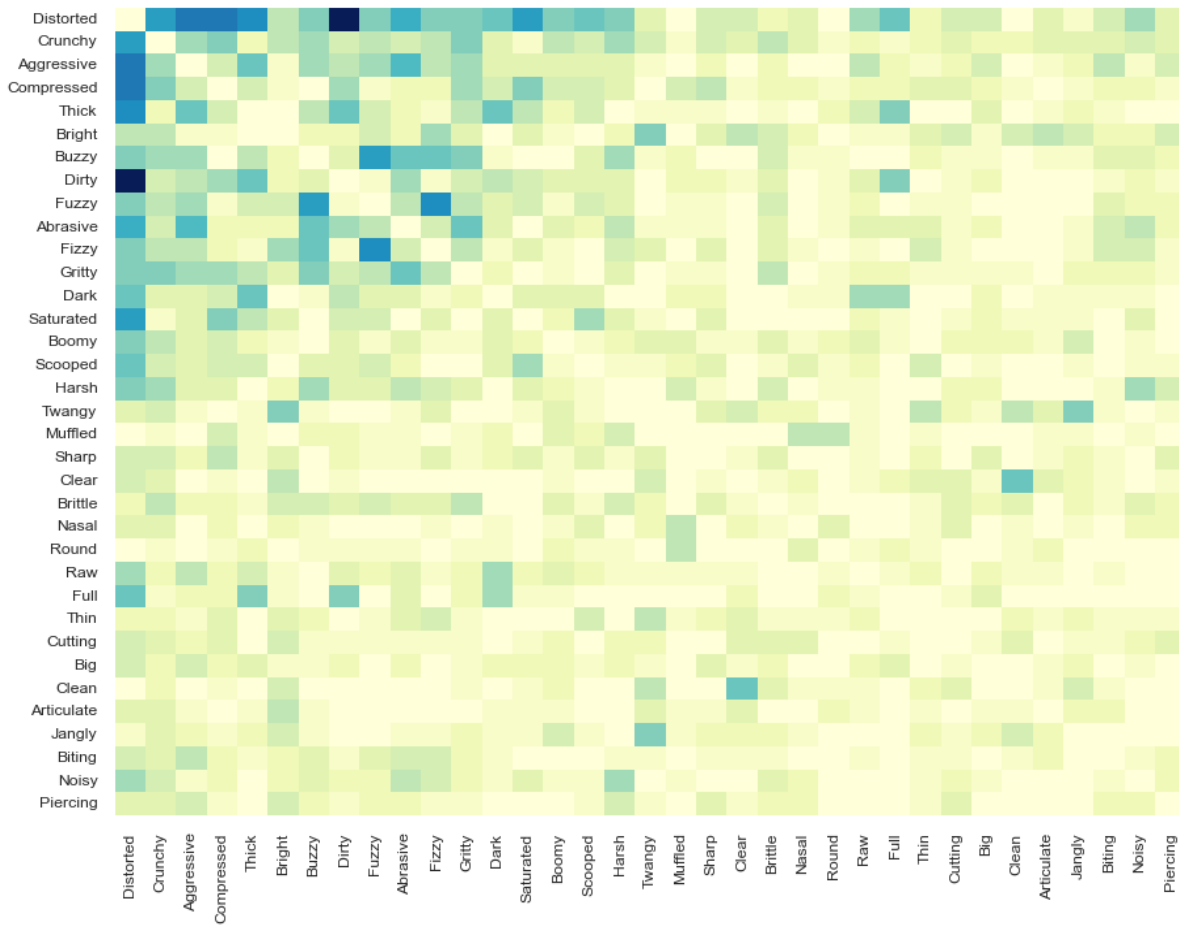

Figure 3: Cross-correlation plot. Darker colors indicate stronger correlations. A win in the rank comparison is treated as a label for that adjective.

## C.3 Cross-Correlation

We also perform a cross-correlation analysis between the clips and adjective labels (the most correlated adjectives are shown in the heatmap in Figure 3). We again observe the most frequent annotations pertaining to heavy or distorted sounds, but we can also observe the extent to which some adjectives may function as synonyms or are otherwise highly correlated. For instance, perhaps unsurprisingly, "*distorted*" and "*dirty*" apply to the same clips. But a "*full*" clip is one that is also "*distorted*" and "*dirty*", but also "*thick*" and often "*dark*". In the absence of additional evidence, this method of defining less understood adjectives in terms of more understood adjectives can help find a more general consensus of meaning for new or unknown words. However, the data can also be used for a more focused study of the audio features based on contrastive examples (for instance, where a recording is labeled as "*thick*" but not "*full*") which can help identify which acoustic properties are most associated with the adjective, and to what extent adjectives are true synonyms.

# D  Limitations

The constructed dataset provides a unique resource for researchers seeking to study the relationship between timbral descriptions and guitar sounds. However, there are limitations to note. Among them, in the era of big data, the number of annotations is relatively small. This is a consequence of the necessity that annotators be experienced in guitar playing and participants in online discussion forums. We present ways of smoothing these statistics to help enable their use in future research, but some estimates may be better represented than others. As there is no objective grounding of these terms, it is difficult to assess the extent to which this is true.

A second concern is that our online approach to data collection allowed users to listen to the clips in their own environments, which may differ significantly from one user to another. However, previous crowdsourcing of timbre descriptions from audio clips have made similar assumptions (Seetharaman and Pardo, 2016). Our addition of pairwise comparison is designed to further mitigate the effect of the environment on labeling, as it establishes a relationship between two recordings.